\documentclass[preprint,aps,prl,showpacs,twocolumn,mamsmath,10pt]{revtex4}
\usepackage{epsfig,epsf,graphics,psfrag}
\usepackage{times}
\usepackage{float}
\usepackage{color}
\usepackage{amsfonts,amssymb,stmaryrd,latexsym,amsmath}
\usepackage{hhline}

\newcommand{\be}{\begin{equation}}
\newcommand{\ee}{\end{equation}}
\newcommand{\ba}{\begin{eqnarray}}
\newcommand{\ea}{\end{eqnarray}}
\newcommand{\non}{\nonumber}

\newcommand{\al}{&\!\!\!}

\begin{document}
\preprint{\small FZJ-IKP-TH-2010-17, HISKP-TH-10/19}
\title{Extracting the light quark mass ratio {\boldmath$m_u/m_d$} from bottomonia transitions}

\author{Feng-Kun Guo$^1$\footnote{Present address: Helmholtz-Institut f\"ur Strahlen- und
             Kernphysik, Universit\"at Bonn,  D--53115 Bonn, Germany}
      , Christoph~Hanhart$^{1,2}$
      , and Ulf-G. Mei{\ss}ner$^{1,2,3}$ 
      }

\affiliation{$\rm ^1$Institut f\"{u}r Kernphysik and J\"ulich Center
             for Hadron Physics, Forschungszentrum J\"{u}lich,
             D--52425 J\"{u}lich, Germany}%
\affiliation{$\rm ^2$Institute for Advanced Simulation,
             Forschungszentrum J\"{u}lich, D--52425 J\"{u}lich, Germany}
\affiliation{$\rm ^3$Helmholtz-Institut f\"ur Strahlen- und
             Kernphysik and Bethe Center for Theoretical Physics,\\ Universit\"at
             Bonn,  D--53115 Bonn, Germany}

\begin{abstract}
\noindent We propose a new method to extract the light quark mass ratio
$m_u/m_d$ using the $\Upsilon(4S)\to h_b\pi^0(\eta)$ bottomonia transitions. The
decay amplitudes are dominated by the light quark mass differences, and the
corrections from other effects are rather small, allowing for a precise
extraction. We also discuss how to reduce the theoretical uncertainty with the
help of future experiments. As a by-product, we show that the decay
$\Upsilon(4S)\to h_b\eta$ is expected to be a nice channel for searching for the
$h_b$ state.
\end{abstract}


\maketitle


Although fundamental parameters of the Standard Model, the masses of light
quarks have not yet been well determined. This appears to be a consequence of
quark confinement as well as the fact that the light quark masses are
significantly lighter than the typical hadronic scale and as such their impact
on most of the hadron masses or other properties is very small.

As a consequence of the spontaneous chiral symmetry breaking, the low-energy
region of the quantum chromodynamics (QCD) can be described by  chiral
perturbation theory (CHPT)~\cite{Weinberg:1978kz,Gasser:1983yg}. The most direct
way to get information on the light quark mass ratios is to relate the quark
masses to the masses of the lowest-lying pseudoscalar mesons, which are the
Goldstone bosons of the spontaneously broken chiral symmetry of QCD. To leading
order (LO) in the chiral expansion, this gives
$m_u/m_d=0.56$~\cite{Weinberg:1977hb}. Electromagnetic (e.m.) effects have been
taken into account using Dashen's theorem~\cite{Dashen:1969eg}. There might be,
however, sizeable higher order corrections to this LO result, e.g. related to
violations of Dashen's theorem, see  \cite{Bijnens:1996kk,Moussallam:1997xx}.
The up-to-date knowledge of the light quark mass ratio from various sources
including recent lattice calculations was summarized in
Ref.~\cite{Leutwyler:Stern} to be
\begin{equation}
\frac{m_u}{m_d}=0.47\pm0.08 \ .
\label{qmr_leutw}
\end{equation}

In a completely independent approach it was proposed to use the decays of
$\psi'$ into $J/\psi\pi^0$ and $J/\psi\eta$, which break isospin and SU(3)
symmetry, respectively~\cite{Ioffe:1979rv,Ioffe:1980mx}. It was assumed that
these decays are dominated by the emission of soft gluons, and the gluons then
hadronize into a pion or an eta. Using the QCD multipole expansion (QCDME), one
obtains
\ba%
\frac{\Gamma(\psi'\to J/\psi\pi^0)}{\Gamma(\psi'\to J/\psi\eta)} = 3
\left(\frac{m_d-m_u}{m_d+m_u}\right)^2 \frac{F_\pi^2}{F_\eta^2}
\frac{M_\pi^4}{M_\eta^4} \left|{\vec{q}_\pi\over\vec{q}_\eta}\right|^3,
\ea%
where $F_{\pi(\eta)}$ and $M_{\pi(\eta)}$ are the decay constant and mass of the
pion (eta), respectively, and $\vec{q}_{\pi(\eta)}$ is the pion (eta) momentum
in the $\psi'$ rest frame. These two decays were  widely used in determining the
quark mass ratio
$m_u/m_d$~\cite{Gasser:1982ap,Donoghue:1992ac,Donoghue:1993ha,Leutwyler:1996qg}.
Using the most recent measurement of the decay widths from the CLEO
Collaboration~\cite{Mendez:2008kb}, one gets $m_u/m_d=0.40\pm0.01$, which is
much smaller than the one resulting from the meson masses. Using instead the
measurement by the BES Collaboration~\cite{Bai:2004cg}, the resulting value
$m_u/m_d=0.35\pm0.02$ is even smaller. 
In Ref.~\cite{Guo:2009wr}, based on a non-relativistic effective field theory
(NREFT) formalism, the striking discrepancy between the values of $m_u/m_d$
extracted from the $\psi'$ decays and from the meson masses was solved by
showing that the decay amplitudes of the transitions $\psi'\to
J/\psi\pi^0(\eta)$ are not dominated by the multipole effect as assumed before.
Rather, non-multipole effects via intermediate charmed meson loops are very
important, enhanced by $1/v$, $v\simeq 0.5$ being the charmed meson velocity,
compared with the multipole one. More precisely, the large uncertainty related
to the non-multipole contributions prevents one from an extraction of $m_u/m_d$
from these decays.

In this Letter, we propose a new way to extract the light quark mass ratio using
the transitions of the excited bottomonium $\Upsilon(4S)$ into $h_b \pi^0$ and
$h_b\eta$. Similar to the transitions between charmonium states, the e.m.
contribution to the isospin breaking decay $\Upsilon(4S)\to h_b \pi^0$ is
negligibly small~\cite{Donoghue:1985vp,Maltman:1990mp,fulllength}. This provides
the possibility of extracting the light quark mass ratio from these decays. It
will be shown that the non-multipole effects from intermediate bottom meson
loops are suppressed, and hence the decay amplitudes are proportional to the
light quark mass differences.

The spin-singlet P-wave bottomonium $h_b$ has not been observed yet, however, it
is expected to agree in mass with the spin-averaged mass of the  spin-triplet
P-wave bottomonia $\chi_{bJ}$ (see, e.g., Ref.~\cite{Godfrey:2002rp}), which is
$M_{h_b} = 9900$~MeV.  The $\Upsilon(4S)$ with a mass of $10579.4\pm1.2$~MeV and
width $20.5\pm2.5$~MeV is the first bottomonium above the $B\bar B$ threshold,
and it decays into $B\bar B$ with more than 96\% branching
fraction~\cite{Amsler:2008zzb}. The mass difference between the $\Upsilon(4S)$
and the $h_b$ is about 680~MeV. Hence, both the transitions $\Upsilon(4S)\to
h_b\pi^0$ and $\Upsilon(4S)\to h_b\eta$ are kinematically allowed.

Let us consider the multipole decay mechanism with the light meson being
directly emitted from the bottomonium first, which is described by a tree-level
diagram based on hadronic degrees of freedom. Because the decays are in an
S-wave and break isospin or SU(3) symmetry, the LO amplitude must scale as the
quark mass difference
\be%
{\cal M}^{\rm tree} \sim \delta,
\ee%
with $\delta=m_d-m_u$ for the transition $\Upsilon(4S)\to h_b\pi^0$ and
$\delta=m_s-\hat{m}$, with $\hat{m}=(m_u+m_d)/2$, for the transition
$\Upsilon(4S)\to h_b\eta$.

Corrections to the tree-level result arise due to intermediate heavy meson loops
and higher order terms in the chiral expansion. The loops can be studied in the
framework of the NREFT because the velocity of the heavy meson in the loops is
small. The value of the bottom meson velocity for the transitions considered
here may be estimated as $v\sim
\sqrt{[2M_{\hat{B}}-(M_{\Upsilon(4S)}+M_{h_b})/2]/M_{\hat{B}}}\simeq 0.3$ with
$M_{\hat{B}}$ the averaged bottom meson mass. This estimate is consistent with
determinations of the bottom quark velocity in bottomonium systems based on
non-relativistic QCD (see, e.g. Ref.~\cite{Lepage:1992tx}). For a transition
between a P-wave and an S-wave heavy quarkonium with the emission of a pion or
an eta, it has been shown that the contribution to the decay amplitude from the
intermediate heavy meson loops scales as~\cite{Guo:2010zk}
\be %
\label{eq:pcloop} {\cal M}^{\rm loop} \sim \frac1{v^3}\frac{\vec{q}\
^2}{M_H^2}\Delta,
\ee%
where $\vec{q}$ is the three-momentum of the light meson in the rest frame of
the decaying heavy quarkonium, $M_H$ is the mass of the intermediate heavy
meson, and the meson mass difference $\Delta$ encodes the violation of the
isospin symmetry for the pionic transition or SU(3) symmetry for the eta
transition. Eq.~(\ref{eq:pcloop}) arises because the non-relativistic loop
integral measure contains three powers of momentum, and scales as $v^3$. After
performing the contour integration of the energy, two propagators are left, and
each of them scales as $1/v^2$. The P-wave coupling of the light meson to the
heavy meson gives a factor of $\vec{q}$. The coupling of the heavy mesons to the
P- and S-wave heavy quarkonia are in S- and P-wave, respectively. The P-wave
vertex provides a momentum in the loop integral, and it must be contracted  with
the external momentum of the light meson $\vec{q}$. So the three vertices
together provide a factor of $\vec{q}\ ^2$. Since we are considering isospin or
SU(3) symmetry breaking transitions, the decay amplitude from the loops
is non-vanishing because the heavy mesons within the same isospin or SU(3)
multiplet have different masses. One may pull out the meson mass difference
explicitly to represent the symmetry breaking. Because it is an energy scale and
should be counted as $v^2$, one needs to divide it by $v^2$ for balance. Putting
all pieces together, one gets $[v^3/(v^2)^2][\vec{q}\ ^2/M_H^2][\Delta/v^2]$,
where $1/M_H^2$ is introduced to match dimensions, and Eq.~(\ref{eq:pcloop})
follows. This kind of non-relativistic power counting has already been confirmed
by explicit calculations of the loops~\cite{Guo:2009wr,Guo:2010zk,fulllength}.

To determine the relative size of the loop amplitude compared to the tree-level one,
in order to find out whether the tree-level contribution is dominant, one
should compare the meson mass difference $\Delta$ and the quark mass difference
$\delta$, and estimate the value of the dimensionless pre-factor $\vec{q}\
^2/(v^3M_H^2)$. The momenta of the pion and eta in the final states of
$\Upsilon(4S)\to h_b\pi^0(\eta)$ are 645~MeV ad 389~MeV, respectively. Taking
$v\approx 0.3$ for the velocity, the dimensionless factor $\vec{q}\
^2/(v^3M_B^2)$ is about 0.6 for the pionic transition and 0.2 for the eta
transition. One cannot naively assign the meson mass differences as the same
order as the quark ones. In fact, due to destructive interference between the
quark mass difference and the e.m. contribution~\cite{Guo:2008ns}, the isospin
mass splitting of the bottom mesons $B^0$ and $B^+$ is rather small,
$M_{B^0}-M_{B^+} = 0.33\pm0.06~{\rm MeV}$~\cite{Amsler:2008zzb}. It is one order
of magnitude smaller than $m_d-m_u$. Together with the dimensionless factor,
which is about 0.6, the bottom meson loops contribute to the decay $\Upsilon(4S)\to
h_b\pi^0$ for no more than a few percent, and hence are negligible. The situation
for the eta transition is somewhat different because
$M_{B_s}-\hat{M}_{B}=87.0\pm0.6$~MeV, where $\hat{M}_{B}=(M_{B^0}+M_{B^+})/2$,
and it is of similar size as $m_s-\hat{m}$. This means that the loop contributions
to the $\Upsilon(4S)\to h_b\eta$ as compared to the tree-level decay amplitude
are also suppressed, but they might give a non-vanishing correction of about 20\%.

An intriguing implication of the suppression of the bottom meson loops in these
transitions is that the decay amplitudes are dominated by the quark mass
differences, and hence it is possible to extract the light quark mass ratio from
the ratio of the branching fractions of the transitions $\Upsilon(4S)\to
h_b\pi^0(\eta)$ with good accuracy. It has been demonstrated that the LO results
of chiral Lagrangians for the heavy quarkonia transitions can reproduce the LO
results of the QCDME~\cite{Casalbuoni:1992fd}. In the QCDME, the transitions
between two heavy quarkonia occur through radiating soft gluons, and the soft
gluons then hadronize into light
mesons~\cite{Gottfried:1977gp,Voloshin:1978hc,Kuang:2006me}. In the case of
transitions with the emission of a pion or an eta, the gluon operator is
$G\tilde{G}\equiv \alpha_s G_{\mu\nu}^A\tilde{G}^{A\mu\nu}$~\cite{Ioffe:1980mx},
where $\alpha_s$ is the strong coupling constant, $G_{\mu\nu}^A$ is the gluon
field strength tensor and its dual is
$\tilde{G}^{A\mu\nu}=\epsilon^{\mu\nu\rho\sigma}G_{\rho\sigma}^A/2$. For the
transitions $\Upsilon(4S)\to h_b\pi^0(\eta)$, we have
\ba%
\label{eq:Rpi0eta} \frac{\Gamma(\Upsilon(4S)\to
h_b\pi^0)}{\Gamma(\Upsilon(4S)\to h_b\eta)} = r_{G\tilde G}^2
\left|{\vec{q}_\pi\over\vec{q}_\eta}\right|,
\ea%
where $\vec{q}_{\pi(\eta)}$ is the momentum of the pion (eta) in the rest frame
of the $\Upsilon(4S)$, and the ratio of the gluon matrix elements is defined as
\be%
r_{G\tilde G} \equiv
\frac{\langle 0| G\tilde{G} |\pi^0\rangle }{ \langle 0|G\tilde{G}|\eta\rangle}~.
\ee%
Combining CHPT with the U(1)$_A$ anomaly, the next-to-leading order (NLO)
expressions for the matrix elements
$\langle 0|G\tilde{G}|\pi^0(\eta)\rangle$ were worked out in
Ref.~\cite{Donoghue:1992ac} in terms of several low-energy constants (LECs) of
the ${\cal O}(p^4)$ Lagrangian. Moreover, there exists an intriguing relation between
the ratio of the matrix elements and a combination of the light quark
masses~\cite{Donoghue:1992ac,Donoghue:1993ha}
\ba%
\label{eq:DWratio} r_{\rm DW} \al\equiv\al \frac{m_d-m_u}{m_d+m_u}
\frac{m_s+\hat{m}}{m_s-\hat{m}} \non\\
\al=\al \frac{4}{3\sqrt{3}} r_{G\tilde G} \frac{F_\pi}{F_\eta}
\frac{F_K^2M_K^2-F_\pi^2M_\pi^2}{F_\pi^2M_\pi^2} (1-\delta_{\rm GMO}) \non\\
\al\al \times \left[1+\frac{4L_{14}}{F_\pi^2} (M_\eta^2-M_\pi^2)\right]
\non\\
\al=\al 10.59 \, (1+132.1 \, L_{14}) \, r_{G\tilde G} ,
\ea%
where $\delta_{\rm GMO}=-0.06$ denotes the ${\cal O}(p^4)$ deviation from the
Gell-Mann--Okubo relation among the Goldstone bosons. Higher order terms in the
coupling of the flavor-singlet field that encodes the information of the
anomalously broken U(1)$_A$ anomaly are parameterized by the ${\cal O}(p^4)$ LEC
$L_{14}$. Therefore, once one has  knowledge of the value of the LEC $L_{14}$,
one is able to extract the quark mass ratio from the ratio of branching
fractions of the decays $\Upsilon(4S)\to h_b\pi^0$ and $\Upsilon(4S)\to
h_b\eta$, which can be measured in the future.

There are two main theoretical uncertainties for extracting the value of $r_{\rm
DW}$. The first one is due to lack of knowledge of the LEC $L_{14}$. One may use
resonance saturation to estimate its value, and it is expected to be in the
region~\cite{Donoghue:1992ac,Donoghue:1993ha} $L_{14}=
(2.3\pm1.1)\times10^{-3}$. From Eq.~(\ref{eq:DWratio}), it gives 11\%
uncertainty in $r_{\rm DW}$. The other one is from neglecting the intermediate
bottom meson loops of the transition $\Upsilon(4S)\to h_b\eta$. As already
discussed, it gives an uncertainty of 20\% in the amplitude, and hence 40\% in
the decay width. Propagating to the extracted quark mass ratio, the uncertainty
is again 20\%. Adding them quadratically, the theoretical uncertainty for
extracting $r_{\rm DW}$ is 23\% which is comparable to that of
Eq.~(\ref{qmr_leutw}).

The uncertainty could be reduced once further information on the size of the
loops is available. This kind of information could be provided by high
statistics measurements in the following way:  The decay width for
$\Upsilon(4S)\to h_b\eta$ considering only loops can be worked out using the
NREFT
\be%
\label{eq:loopwidth} \Gamma(\Upsilon(4S)\to h_b\eta)^{\rm loop} = 0.16
g_{1b}^2~{\rm keV},
\ee%
where the only unknown parameter $g_{1b}$ denotes the coupling of the $1P$
bottomonium states to the bottom mesons, given in GeV$^{-1/2}$. Although the
$1P$ states are below any open bottom threshold, one may extract $g_{1b}$ from
the {\em loop dominated transitions} involving the $1P$ states. Because the
isospin mass splitting of the $B$ mesons is rather small, one should consider
the loop dominated transitions with the emission of an $\eta$. These are the
transitions from excited P-wave bottomonia to the $1P$ states, enhanced by a
factor $1/v^3$, using a similar power counting technique presented in
Ref.~\cite{fulllength}. The best choice are the $\eta$ transitions from the $4P$
states to the $1P$ states. Based on the quark model calculation of the
bottomonium spectrum~\cite{Li:2009nr}, the $4P$ states have sufficiently large
masses to allow for decays into $B^{(*)}\bar{B}^{(*)}$. Because the $4P$ states
can decay directly into $B^{(*)}\bar{B}^{(*)}$, the corresponding coupling
constant $g_{1b}'''$ can be obtained by measuring their decay widths. Then, one
can extract the value of $g_{1b}$ from any of the transitions $\chi_{b0}(4P)\to
\chi_{b1}\eta$, $\chi_{b1}(4P)\to \chi_{b0,1,2}\eta$, $\chi_{b2}(4P)\to
\chi_{b1,2}\eta$ and $h_b(4P)\to h_b\eta$.
\begin{figure*}[t]
\begin{center}
\vglue-0mm
\includegraphics[width=0.32\textwidth]{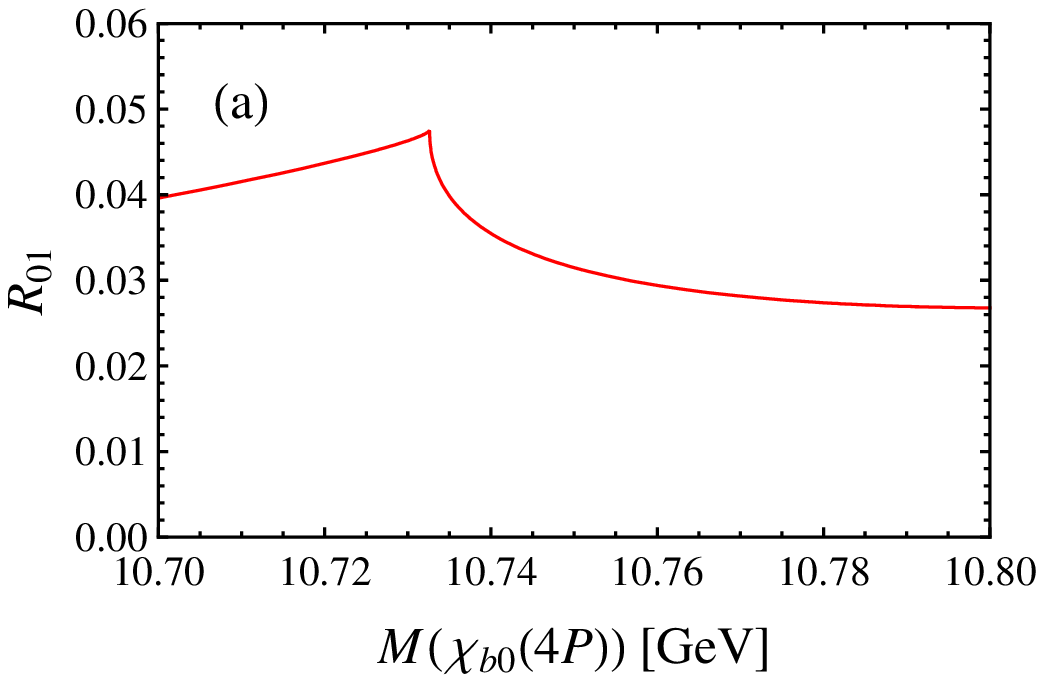}\hfill
\includegraphics[width=0.32\textwidth]{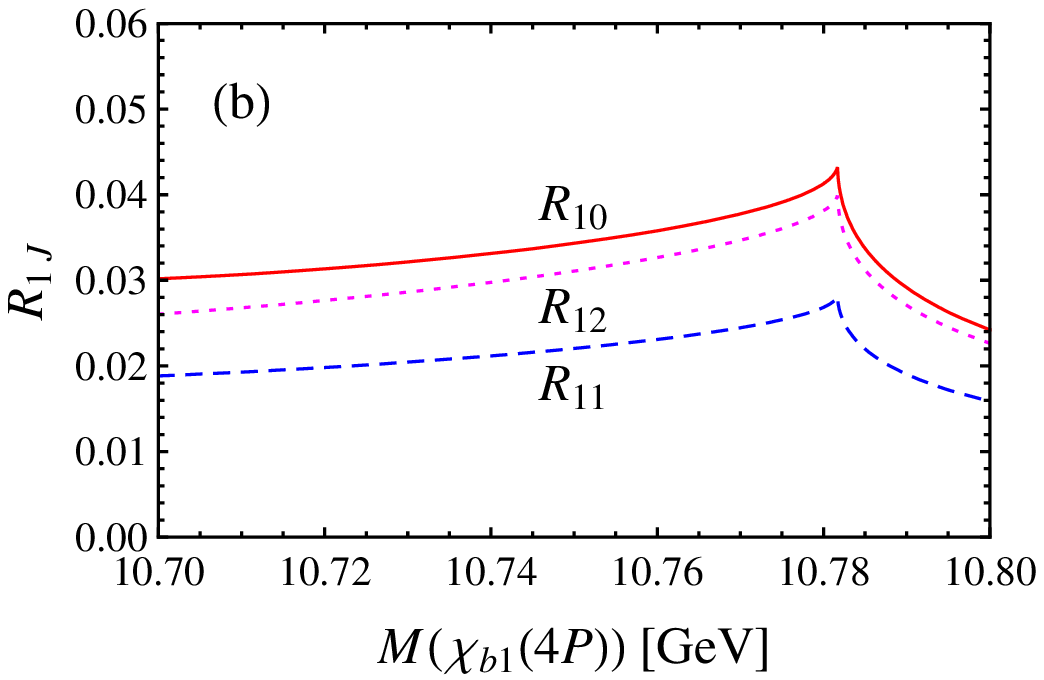}\hfill
\includegraphics[width=0.32\textwidth]{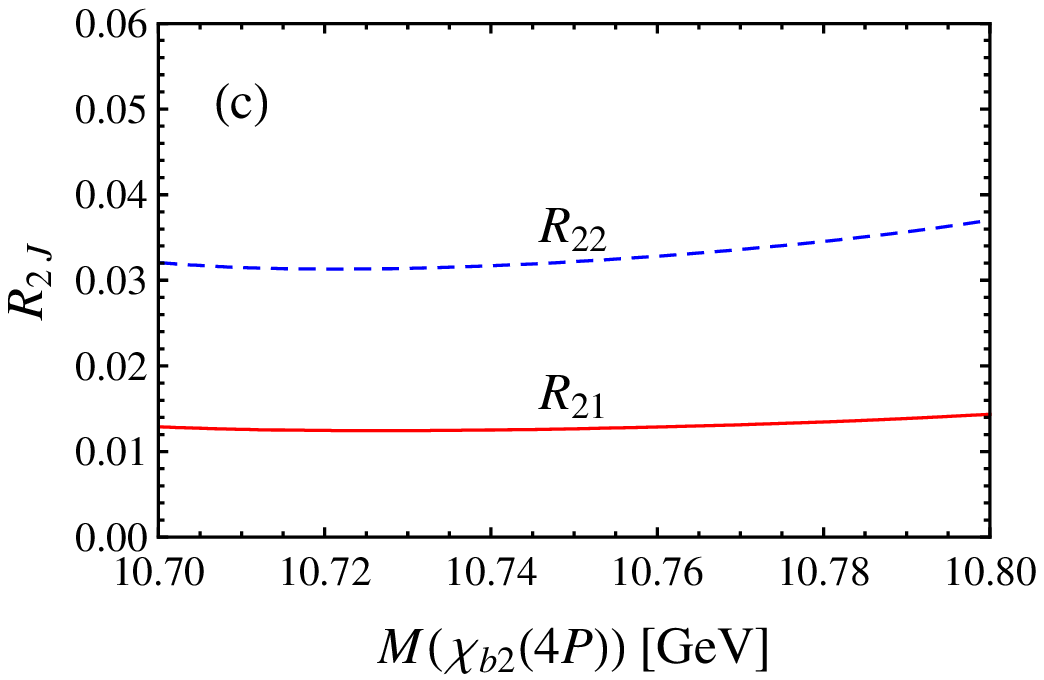}
\vglue-0mm \caption{Predicted ratios defined in Eq.~(\ref{eq:ratios}) in the
NREFT. The value of $g_{1b}$ has been set to 1~GeV$^{-1/2}$. \label{fig:g1b}}
\end{center}
\end{figure*}
In Fig.~\ref{fig:g1b}, we show the predictions from the NREFT for the following
ratios, which depend on $g_{1b}$ only and are {\em proportional to} $g_{1b}^2$,
as a function of the mass of the $4P$ bottomonium state:
\ba%
R_{01} \al\equiv\al
\frac{\Gamma(\chi_{b0}(4P)\to\chi_{b1}\eta)}{\Gamma(\chi_{b0}(4P)\to B^+B^-)},\non\\
R_{1J} \al\equiv\al
\frac{\Gamma(\chi_{b1}(4P)\to\chi_{bJ}\eta)}{\Gamma(\chi_{b1}(4P)\to
B^+B^{*-})}, \quad J=0,1,2,
\non\\
R_{2J} \al\equiv\al
\frac{\Gamma(\chi_{b2}(4P)\to\chi_{bJ}\eta)}{\Gamma(\chi_{b2}(4P)\to
B^{*+}B^{*-})}, \quad J=1,2. 
\label{eq:ratios}
\ea%
The result for $\Gamma(h_{b}(4P)\to h_{b}\eta)/\Gamma(h_{b}(4P)\to
B^{*+}B^{*-})$ is very similar to and slightly larger than $R_{10}$. The cusps
in Fig.~\ref{fig:g1b} (a) and (b) represent the opening of the $B_s{\bar B}_s$
and $B_s{\bar B}_s^*$ thresholds, respectively. For definiteness, we have used
$g_{1b}=1$~GeV$^{-1/2}$. The dependence on $g_{1b}'''$ is canceled in the
ratios. If any of these ratios were to be measured, one will be able to extract
the value of $g_{1b}^2$ easily.
The so extracted coupling $g_{1b}^2$ bears about 30\% uncertainty due to the
loops of higher order. The $4P$ bottomonia should decay dominantly into
$B^{(*)}{\bar B}^{(*)}$. For an order-of-magnitude estimate, one may assume
$g_{1b}$ to have a similar value as its charm analogue estimated using vector
meson dominance, $g_{1c}=-4.2$~GeV$^{-1/2}$~\cite{Colangelo:2003sa}. Then from
Fig.~\ref{fig:g1b}, one expects the eta transitions have branching fractions of
the order of a few per cent. Hence, there should be a good opportunity in
extracting $g_{1b}$ at the Large Hadron Collider beauty experiment (LHCb).
After having measured the partial decay width of $\Upsilon(4S)\to h_b\eta$, one
may compare the measured value with the one obtained considering only the bottom
meson loops, given in Eq.~(\ref{eq:loopwidth}), using $g_{1b}$ determined in the
way outlined above as input. Depending on whether the interference between the
tree-level and the loop amplitudes is constructive or destructive, one can get
two solutions of the width considering only the multipole (tree-level) effect.
Then one may insert the resulting $\Gamma(\Upsilon(4S)\to h_b\eta)^{\rm tree}$
in Eq.~(\ref{eq:Rpi0eta}), reducing the uncertainty from the loop contribution.

Using the same naturalness arguments for the coupling constant in the LO
tree-level Lagrangian as that in Ref.~\cite{Guo:2010zk}, we can estimate the
branching fractions of the transitions $\Upsilon(4S)\to h_b\pi^0(\eta)$. The
branching fraction for the pionic transition is of order $10^{-6}$, and the one
for the eta transition is of order $10^{-3}$. With such a large branching
fraction, the latter one even provides a nice option for searching for the $h_b$.
LHCb is expected to have enough events of the $\Upsilon(4S)$ to do the measurements.

In summary, we have proposed a new method for extracting  $m_u/m_d$. We
demonstrated that the transitions $\Upsilon(4S)\to h_b\pi^0(\eta)$ can be used
to determine the value of $r_{\rm DW}$ with an acceptable theoretical
uncertainty, which is about 23\%. Using information of $\hat{m}/m_s$ from other
sources, one is then able to extract $m_u/m_d$.
The transitions $\Upsilon(4S)\to h_b\pi^0$ and $\Upsilon(4S)\to h_b\eta$ are
expected to have branching fractions of order of $10^{-6}$ and $10^{-3}$,
respectively. Therefore, they can be measured at LHCb based on a large number of
$\Upsilon(4S)$ events. The uncertainty can be reduced to obtain a more accurate
extraction of the quark mass ratio by measuring the partial decay widths of the
$4P$ bottomonium to the $1P$ bottomonium with the emission of an eta. These
transitions with branching fractions of order of a few per cent can also be
measured at LHCb. As a by-product, the decay $\Upsilon(4S)\to h_b\eta$ is
expected to be a nice channel for searching for the $h_b$ state.


This work is partially supported by  the Helm\-holtz Association through funds
provided to the Virtual Institute ``Spin and strong QCD'' (VH-VI-231) and by the
DFG (TR 16, ``Subnuclear Structure of Matter''), and the European
Community-Research Infrastructure Integrating Activity ``Study of Strongly
Interacting Matter'' (acronym HadronPhysics2, Grant Agreement n. 227431) under
the FP7 of EU, and by BMBF (Grant No. 06BN9006).


\begin{thebibliography}{99}

\bibitem{Weinberg:1978kz}
  S.~Weinberg,
  Physica A {\bf 96}, 327 (1979).

\bibitem{Gasser:1983yg}
  J.~Gasser and H.~Leutwyler,
  Annals Phys.\  {\bf 158}, 142 (1984);
  Nucl.\ Phys.\  B {\bf 250}, 465 (1985).


\bibitem{Weinberg:1977hb}
  S.~Weinberg,
  in {\it A Festschrift for I.I. Rabi}, ed. by L. Motz [Trans.\ New York Acad.\ Sci.\  {\bf 38}, 185 (1977)].

\bibitem{Dashen:1969eg}
  R.~F.~Dashen,
  Phys.\ Rev.\  {\bf 183}, 1245 (1969).

%
\bibitem{Bijnens:1996kk}
  J.~Bijnens and J.~Prades,
  Nucl.\ Phys.\  B {\bf 490}, 239 (1997).

\bibitem{Moussallam:1997xx}
  B.~Moussallam,
  Nucl.\ Phys.\  B {\bf 504}, 381 (1997).

\bibitem{Leutwyler:Stern}
  H.~Leutwyler,
  in Colloquium in memory of Jan Stern, Paris, France, 2009 (unpublished).

\bibitem{Ioffe:1979rv}
  B.~L.~Ioffe,
  Yad.\ Fiz.\  {\bf 29}, 1611 (1979) [Sov. J. Nucl. Phys. {\bf 19}, 827 (1979)].

\bibitem{Ioffe:1980mx}
  B.~L.~Ioffe and M.~A.~Shifman,
  Phys.\ Lett.\  B {\bf 95}, 99 (1980).

\bibitem{Gasser:1982ap}
  J.~Gasser and H.~Leutwyler,
  Phys.\ Rept.\  {\bf 87}, 77 (1982).

\bibitem{Donoghue:1993ha}
  J.~F.~Donoghue, B.~R.~Holstein and D.~Wyler,
  Phys.\ Rev.\ Lett.\  {\bf 69}, 3444 (1992).

\bibitem{Leutwyler:1996qg}
  H.~Leutwyler,
  Phys.\ Lett.\  B {\bf 378}, 313 (1996).

\bibitem{Donoghue:1992ac}
  J.~F.~Donoghue and D.~Wyler,
  Phys.\ Rev.\  D {\bf 45}, 892 (1992).

\bibitem{Mendez:2008kb}
  H.~Mendez {\it et al.}  [CLEO Collaboration],
  Phys.\ Rev.\  D {\bf 78}, 011102 (2008).

\bibitem{Bai:2004cg}
  J.~Z.~Bai {\it et al.}  [BES Collaboration],
  Phys.\ Rev.\  D {\bf 70}, 012006 (2004).

\bibitem{Guo:2009wr}
  F.-K.~Guo, C.~Hanhart and U.-G.~Mei{\ss}ner,
  Phys.\ Rev.\ Lett.\  {\bf 103}, 082003 (2009)
  [Erratum, {\it ibid} {\bf 104}, 109901 (2010)].

\bibitem{Donoghue:1985vp}
  J.~F.~Donoghue and S.~F.~Tuan,
  Phys.\ Lett.\  B {\bf 164}, 401 (1985).

\bibitem{Maltman:1990mp}
  K.~Maltman,
  Phys.\ Rev.\  D {\bf 44}, 751 (1991).



\bibitem{fulllength}
  F.-K.~Guo, C.~Hanhart, G.~Li, U.-G.~Mei{\ss}ner and Q.~Zhao,
  arXiv:1008.3632 [hep-ph].


\bibitem{Godfrey:2002rp}
  S.~Godfrey and J.~L.~Rosner,
  Phys.\ Rev.\  D {\bf 66}, 014012 (2002).

\bibitem{Amsler:2008zzb}
  C.~Amsler {\it et al.}  [Particle Data Group],
  Phys.\ Lett.\  B {\bf 667}, 1 (2008) and 2009 partial update for the 2010 edition.

\bibitem{Lepage:1992tx}
  G.~P.~Lepage, L.~Magnea, C.~Nakhleh, U.~Magnea and K.~Hornbostel,
  Phys.\ Rev.\  D {\bf 46}, 4052 (1992).

\bibitem{Guo:2010zk}
  F.-K.~Guo, C.~Hanhart, G.~Li, U.-G.~Mei{\ss}ner and Q.~Zhao,
  Phys.\ Rev.\  D {\bf 82}, 034025 (2010).



\bibitem{Guo:2008ns} F.-K.~Guo, C.~Hanhart and U.-G.~Mei{\ss}ner,
  JHEP {\bf 0809}, 136 (2008).

\bibitem{Casalbuoni:1992fd}
  R.~Casalbuoni, A.~Deandrea, N.~Di Bartolomeo, R.~Gatto, F.~Feruglio and G.~Nardulli,
  Phys.\ Lett.\  B {\bf 309}, 163 (1993).

\bibitem{Gottfried:1977gp}
  K.~Gottfried,
  Phys.\ Rev.\ Lett.\  {\bf 40}, 598 (1978).

\bibitem{Voloshin:1978hc}
  M.~B.~Voloshin,
  Nucl.\ Phys.\  B {\bf 154}, 365 (1979).

\bibitem{Kuang:2006me}
  Y.~P.~Kuang,
  Front.\ Phys.\ China {\bf 1}, 19 (2006).

\bibitem{Li:2009nr}
  B.~Q.~Li and K.~T.~Chao,
  Commun.\ Theor.\ Phys.\  {\bf 52}, 653 (2009).

\bibitem{Colangelo:2003sa}
  P.~Colangelo, F.~De Fazio and T.~N.~Pham,
  Phys.\ Rev.\  D {\bf 69}, 054023 (2004).


\end{thebibliography}
\end{document}